# Multi hole bands and quasi-two-dimensionality in $Cr_2Ge_2Te_6$ studied by angle-resolved photoemission spectroscopy

T. Yilmaz[1], R. M. Geilhufe[2], I. Pletikosić[3], G. W. Fernando[4], R. J. Cava[5], T. Valla[6], E. Vescovo[7], and B. Sinkovic[8]

[1]National Synchrotron Light Source II, Brookhaven National Lab, Upton, New York 11973, USA
[2]Nordita, Roslagstullsbacken 23 SE-106 91 Stockholm Sweden
[3]Department of Physics, Princeton University, Princeton, New Jersey 08544, USA
[4]Department of Physics, University of Connecticut, Storrs, Connecticut 06269, USA
[5]Department of Chemistry, Princeton University, Princeton, New Jersey 08544, USA
[6]Condensed Matter Physics and Materials Science Department, Brookhaven National Laboratory, Upton, New York 11973, USA

**In the present work, we investigate the electronic structure of the two-dimensional ferromagnet $Cr_2Ge_2Te_6$ by photoemission spectroscopy and Ab Initio calculations. Our results demonstrate the presence of multiple hole-type bands in the vicinity of the Fermi level indicating that the material can support high electrical conductivity by manipulating the chemical potential. Also, our photon energy-dependent angle resolved photoemission experiment shows that several of the hole bands exhibit weak dispersion with varied incident photon energy providing experimental signature for two dimensional nature of $Cr_2Ge_2Te_6$. Thereby, these findings can pave the way for further studies towards the application of $Cr_2Ge_2Te_6$ in electronic devices.**

## 1 Introduction

Low dimensional quantum materials have surprised the scientific community with their unexpected properties such as strong room-temperature ferromagnetism in monolayer VSe2 and high-temperature superconductivity in monolayer FeSe [1- 3]. Among of many two-dimensional systems, van derWaals (vdW) materials with long-range ferromagnetic order are of particular interest for their potential application in the spin-based electronic devices [4- 7]. The materials $Cr_2Ge_2Te_6$ is one of the few vdW systems with ferromagnetic order while being a semiconductor [8]. Along with its rich optical and electronic properties, $Cr_2Ge_2Te_6$ can also promote various experimental platforms such as magnetizing the surface states of a topological insulator through the growth of a heterostructure to realize the anomalous quantum Hall effect [9]. In a recent work, $Cr_2Ge_2Te_6$ is shown to host the ferromagnetic state down to a monolayer with ferromagnetic transition temperature (Tc) of 106 K while the bulk $Cr_2Ge_2Te_6$ has $T_c$ of 61 K [4, 10]. Furthermore, $Cr_2Ge_2Te_6$ is also proposed as a promising thermoelectric material which can convert heat into electricity for use in daily life [11-12]. However, the power factor of $Cr_2Ge_2Te_6$ is as small as 0.23 mW/mK$^2$ based on recent findings [12]. But it was suggested that Mn doping into the bulk can increase the number of the hole bands crossing the Fermi level (EF) and enhances the power factor from 0.23 mW/mK$^2$ to 0.57 mW/mK$^2$ at 830 K [12].

Even though the band structure of $Cr_2Ge_2Te_6$ was well studied by theory [12-15], the experimental works are limited to a few to the best of our knowledge [16- 18]. Most importantly, its quasi-two-dimensional nature is not well studied by photoemission spectroscopy, yet. Here, we report an angle resolved photoemission spectroscopy (ARPES) experiment performed on a $Cr_2Ge_2Te_6$ single crystal by using synchrotron radiation. We found that the vicinity of the EF is dominated by two sets of hole bands with the same band symmetry. By tuning the chemical potential of the system, the number of hole bands crossing EF can be increased. In addition, we show two low lying energy bands near the EF that show a weak dispersion along the $k_z$ momentum direction of the Brillouin zone providing experimental evidence for the quasi-two-dimensional nature of $Cr_2Ge_2Te_6$. We support our experimental findings by further performing Ab Initio calculations.

## 2 Methods

Cr (99.99%), Ge (99.999%), and Te (99.999%) were employed, mixed in a molar ratio of 2 : 6 : 36; the extra Ge and Te were used as a flux. The materials were heated to 700 C, held for 20 days, and then slowly cooled to 500 C over a period of 1.5 days. This was followed by centrifugation to remove the flux. The resultant platelet single crystals are a few millimeters in size. The additional details about the structural properties can be found in Ref. 8. The ARPES experiments were performed at the beamline U5UA at the National Synchrotron Light Source. Core level and polarization-dependent ARPES experiments were carried out at the 21-ID-1 beamline of NSLS-II. Photon energy is converted into $k_z$ momentum by using the free-electron final state approximation, $\hbar k_z = 2m_e\sqrt{E_{kin} \cos^2\theta + V_o}$ where $m_e$ is the free electron mass, $E_{kin}$ is the kinetic energy of a photoelectron, and $V$ is the inner potential taken as 10 eV [16]. Additionally, we have performed first-principles calculations, both on a full trigonal unit cell (space group R3, #148) as well as on a reduced triclinic cell. The calculations were performed within the framework of the density functional theory [19, 20, 21], as implemented in the Vienna Ab Initio Simulation Package (VASP) [22] and the Quantum Espresso code [23]. The exchange-correlation functional was approximated by the generalized gradient approximation [24]. The energy cutoff is chosen according to potential input files. For integration in $\vec{k}$ - space, a 6 x 6 x 2 (trigonal cell) and a 6 x 6 x 6 (triclinic cell) Γ centered mesh according to Monkhorst and Pack [25] was used during the self-consistent cycle. Structural optimization was performed until the forces acting on the atoms were negligible.

## 3 Results and Discussion

Local chemical environment of the $Cr_2Ge_2Te_6$ sample is investigated by high-resolution core-level spectroscopy performed with a synchrotron radiation. Fig. 1(a) shows photoemission intensity versus binding energy region covering the Cr 2p and Te 4d core levels. Both peaks exhibit the same spin orbit splitting energy of 10.4 eV between the Te $4d_{5/2}$ - $4d_{3/2}$ and Cr $2p_{3/2}$ - $3p_{1/2}$ core levels. The Cr $2p_{3/2}$ and $2p_{1/2}$ peaks are located at 576.2 eV and 582.6 eV binding energies whereas the Te $4d_{5/2}$ - $4d_{3/2}$ peaks have the binding energies of 582.6 eV and 572.2 eV, respectively. These binding energies correspond to 3+ oxidation state for Cr and 2- for Te [26, 27]. In addition, the Ge $3p_{3/2}$ and $3p_{1/2}$ peaks in Fig. 1(b) have the binding energies of 127.8 eV and 123.6 eV with a 4.2 eV spin orbit splitting energy. These results show that compared to metallic forms, the Te 4d peaks shifts 0.9 eV towards lower binding energy, while the Cr 3p and the Ge 3p peaks exhibit 1.8 eV and 1.4 eV shifts to higher binding energies, respectively [28].

To explore the electronic structure of $Cr_2Ge_2Te_6$, we present ARPES maps in Fig. 2(a) obtained from a vacuum cleaved sample. The electronic structure was recorded along the G - K direction with hv = 20 eV corresponding to $k_z$ = 2.585 Å$^{-1}$ in the Brillouin zone. The ARPES map exhibits two sets of hole bands labeled with H1 and H2 dominating the vicinity of the E$_F$. These bands are centered at k$_k$ = 0 Å$^{-1}$ (Γ). Another band labeled H3 in Fig. 2(b) is also observed at higher binding energies overlapping with the H2 band at 0.4 eV binding energy. Based on the recent theoretical findings in Ref. 11 and Ref. 14, H1 and H2 bands are formed dominantly by Te 6p and with the non-negligible contribution from Cr 3d atomic orbitals. In addition, the ferromagnetism in Cr$_2$Ge$_2$Te$_6$ is induced by the superexchange mechanism between filled the Cr t$_{2g}$ and the empty Cr e$_g$ states through the p orbitals of the Te atoms [14]. Furthermore, we present the constant energy contours at E$_F$ and at -0.6 eV in Fig. 2(b-c), respectively. The Fermi surface consists of a circularly shaped spectral weight at the Brillouin zone center dominated by the H1 hole band. Away from the E$_F$, H1 and H2 sets can be distinguished (Fig. 2(b)) marked with red and black dashed circles. It is also clearly seen that the H3 band labeled with a yellow dashed line in Fig 2c exhibits a parabolic momentum dispersion that intersects with the momentum contours formed by the H1 band.

To study the two-dimensional nature of Cr$_2$Ge$_2$Te$_6$, we investigate the electronic structure in further detail by performing computational band structure analysis and a photon energy dependent ARPES experiment. The computed band structures with and without spin orbit coupling are given in Fig. 3(a-b), respectively for various high symmetry directions. Even though spin-orbit coupling does not change the overall appearance of the electronic states closest to the E$_F$, its effect on the overall band structure is reasonably strong. Also, the size of the indirect band gap is reduced from about 0.4 eV to about 0.12 eV. Furthermore, below the Fermi level, we find 6 bands which are close in energy and pairwise degenerate. Two of the pairs belong to the mutually complex conjugate one-dimensional irreducible representations $e_u$ and $e_u^*$ whereas the third pair transforms as the mutually complex conjugate one-dimensional irreducible representations $e_g$ and $e_g^*$. The states arise from the Cr d states and the Te p states which get split according to $d \rightarrow a_g \oplus 2e_g \oplus 2e_g^*$ and $p \rightarrow a_u \oplus 2e_u \oplus e_u^*$ in C$_{3i}$ (S$_6$) symmetry [29, 30]. Furthermore, it can be verified from the very flat dispersion along the path G - A of the bands right below the E$_F$, the corresponding Bloch states are quasi-two-dimensional near the vicinity of the G point. To support our theoretical findings, we have conducted a set of ARPES experiments with varied incident photon energies, hence, recording the experimental band structure at different $k_z$ momentum in the Brillouin zone. This approach provides us to construct the band structure along the path Γ – A as presented in Fig. 3(c). The data shows that two sets of bands separated with 0.4 eV energy and marked with dashed red lines in Fig. 3(c) exhibit weak dispersion as a function of $k_z$. The energy separation between two sets of bands and the nearly flat features are in good agreement with our calculations. This provides a spectroscopic evidence for the quasi-two-dimensional nature of Cr$_2$Ge$_2$Te$_6$. In particular, the nondispersive band located in the
vicinity of 0.1 eV binding energy can be seen in the individual ARPES maps taken with 13 eV ($k_z$ = 2.20 Å$^{-1}$) and 19.5 eV ($k_z$ = 2.56 Å$^{-1}$) photon energies (Fig. 3(d-e)). Furthermore, at higher $k_z$, H2 dispersive band can be also seen in the $k_z$ map (Fig. 3(c)) and ARPES maps (Fig. 3(d-e)). Hence, these theoretical and experimental observations verify the occurrence of the quasi-two-dimensional bands along with the dispersive ones in the surface electronic structure of Cr$_2$Ge$_2$Te$_6$. To investigate the orbital character of the hole bands in Cr$_2$Ge$_2$Te$_6$, we compute the partial density of states (DOS) in the vicinity of the E$_F$. As is shown in Fig. 4(a), the valence band is mainly formed by the Te-5p orbitals whereas the conduction band is mixture of the Cr-3d and Te-5p orbitals. In-plane and out-of-plane orbital texture of these states can be further studied experimentally through the light polarization dependence of the photoemission process. ARPES intensity is proportional to the transition matrix element of <f| **E·r** |i> where <f|, |i>, and E are final and initial states of the photoexcited electrons, the electric field vector, and momentum operator, respectively [31]. This will give a non-zero photoemission intensity if the electronic state and **E·r** have the same symmetry[32]. Hence, to study the orbital texture of the hole bands in $Cr_2Ge_2Te_6$, we perform an ARPES experiment with linear horizontal (LH) and linear vertical (LV) polarized 50 eV photons (Fig. 4(b-c)). In the LH geometry, the ARPES map exhibits a strong spectral weight contributed by the H1 and H2 hole bands. On the other hand, the photoemission intensity nearly vanishes for LV geometry. This concludes that the hole bands have the same symmetry since they exhibit the same polarization dependence. This is also a signature of the electronic state of the $p_z$ character in $Cr_2Ge_2Te_6$ as the strong spectral weight suppression with the switching the light polarization from LH to LV .

In summary, we have provided a spectroscopic study together with Ab Initio calculations to demonstrate multiple hole band sets in $Cr_2Ge_2Te_6$. If the $E_F$ can be tuned to higher binding energy, the number of hole bands would increase. This can enhance the thermoelectric property of $Cr_2Ge_2Te_6$. Furthermore, our photon energy-dependent ARPES study confirmed the quasi-two-dimensional nature of $Cr_2Ge_2Te_6$. These results could guide future studies concentrating on the applications of the two-dimensional ferromagnetic materials on information technologies. This research used U5 beamline of the National Synchrotron Light Source I and 21-ID-I beamline of the National Synchrotron Light Source II, U.S. Department of Energy (DOE) Office of Science User Facilities operated for the DOE Office of Science by Brookhaven National Laboratory. We also acknowledge the computing resources provided by the Center for Functional Nanomaterials, which is a U.S. DOE Office of Science Facility, at Brookhaven National Laboratory under Contract No. DE-SC0012704 and support from a Los Alamos grant through the U.S. DOE. This research also used funding from the VILLUM FONDEN via the Centre of Excellence for Dirac Materials (Grant No. 11744), the Knut and Alice Wallenberg foundation (2019.0068) as well as computational resources from the Swedish National Infrastructure for Computing (SNIC).

**Figure 1:** (a-b) High-resolution core level spectra of Cr 2p, Te 3d, and Ge 3p. Red solid lines in (a) and (b) are the overall fitting lines obtained after Shirley type background is subtracted from the spectra. Green, black, blue, cyan, gray, and pink lines are the Voight fitting profiles. Spectra are taken with 800 eV photon energies at 10 K from a vacuum cleaved $Cr_2Ge_2Te_6$ sample.

**Figure 2:** (a) ARPES energy map of the $Cr_2Ge_2Te_6$ sample recorded with hv = 20 eV at the room temperature. (b-c) Constant energy contours at the $E_F$ and -0.6 eV binding energy, respectively. Dashed black, red, and yellow lines in b are for guiding the eyes. Momentum contours formed by H1, H2, and H3 are indicated with dashed red circle, dashed black circle, and dashed yellow parabola, respectively.

**Figure 3:** Band structure calculation for $Cr_2Ge_2Te_6$ in the full trigonal cell: (a) with and (b) without spin orbit coupling. (c) $k_z$ versus binding energy map. (d-e) ARPES maps taken with hv = 13 eV and hv = 19.5 eV, respectively. Dashed black and red lines in c-d are to guide the eyes. ARPES maps were taken at the room-temperature.

**Figure 4:** (a) Calculated partial DOS of $Cr_2Ge_2Te_6$. (b-c) Polarization-dependent ARPES mas of the $Cr_2Ge_2Te_6$ sample taken at 10 K with LH and LV 50 eV lights, respectively. The angle between the incident light and the surface normal to the sample is $55_0$ for the normal emission geometry. ARPES maps were taken at 10 K.

**Figures:**

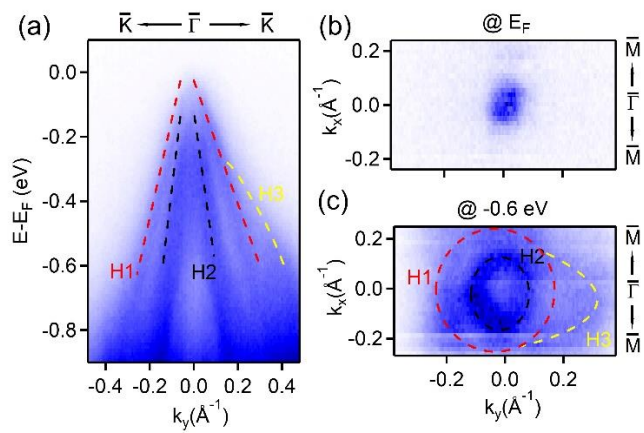

**Figure 1**

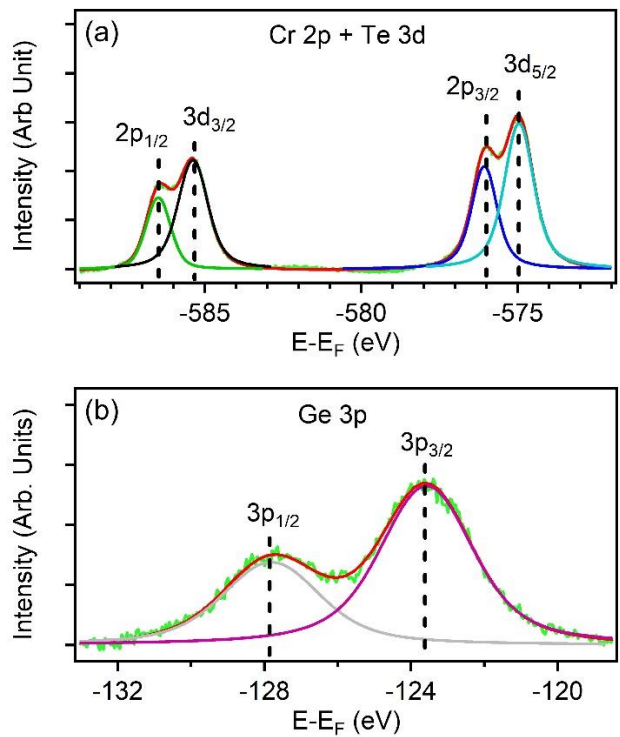

**Figure 2**

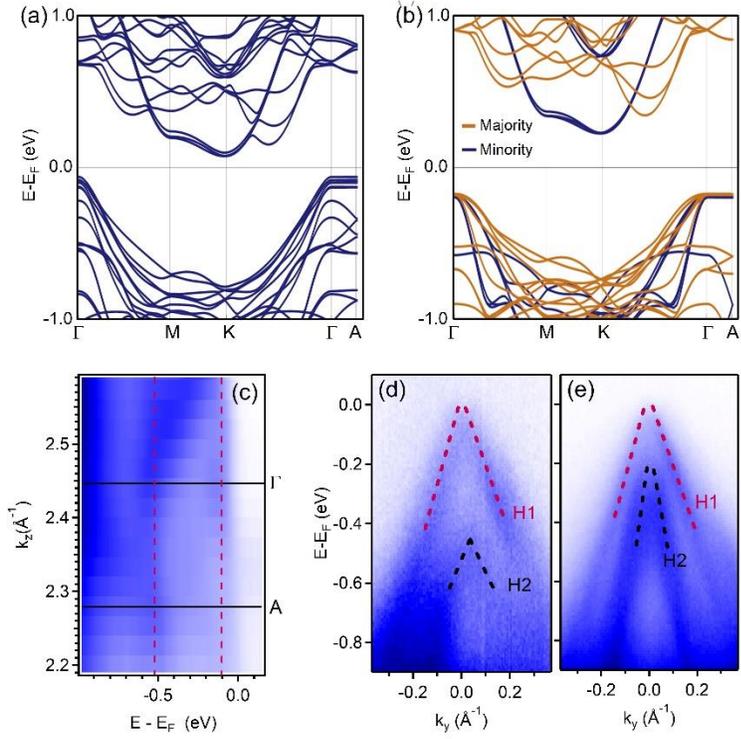

Figure 3

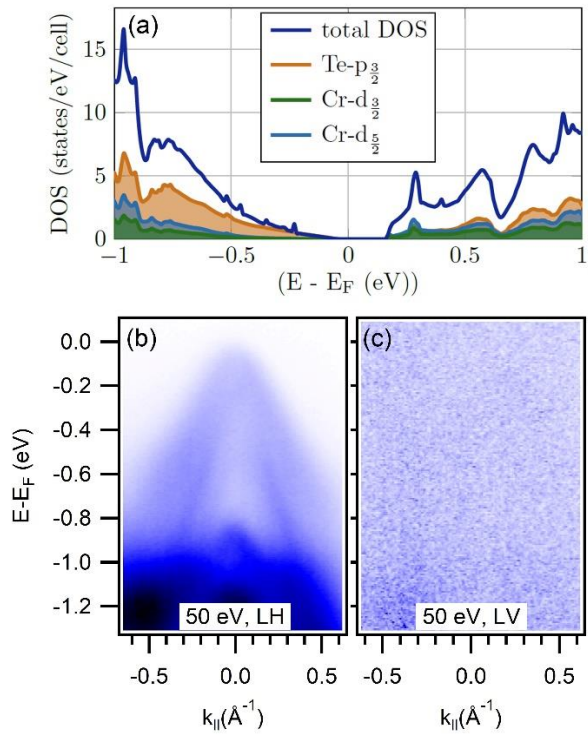

Figure 4